\documentclass{article}
\usepackage{epsfig}
\usepackage{graphicx,caption}

\begin{document}

\centerline{\bf Growth of the inner core in the mean-field dynamo model}

\begin{center}{M. Yu. Reshetnyak  \\ 
\vskip 0.5cm
Institute of Physics of the Earth RAS \\
  Moscow, Russia, m.reshetnyak@gmail.com}
\end{center}

       \abstract{\noindent Application of Parker's dynamo model to the geodynamo with the growing inner core is considered. It is shown that decrease of the inner core size, where intensive magnetic field generation takes place, leads to the multi-polar magnetic field in the past. This effect reflects the decrease of the region of the effective magnetic field generation. The process is accompanied by  increase of the  reversals number and decrease of intensity of the geomagnetic field. The constraints on the mechanisms of convection in the liquid core are discussed.

      \vskip 0.1cm}

 \section{Introduction}  
  Geomagnetic field, generated by the dynamo process in the liquid core of the Earth, is a unique   source of information on the internal structure of the planet. Due to low conductivity of the mantle, magnetic field penetrates from the surface of the inner core to the surface of the planet without significant distortions. The age of the magnetic field, estimated as 3.5Gy  \cite{Tarduno}, see also  
 review \cite{RP}, is compared to the age of the Earth itself 4.5Gy.
       So far during the ``magnetic'' epoch  the liquid and solid cores of the Earth evolved, it is tempting to detect variations of the geomagnetic field, concerned with the evolution of the planet. The main reason of such variations is the formation of the inner core, which effects the dynamo process at least in two ways.

    Firstly, the  growth of the inner core can influence on the magnetic field generation through  the pure geometrical factor, which has no relation  to  the physical mechanisms of convection. The growth of the inner core leads to decrease of the liquid core, and to  increase of the radius of the Taylor cylinder (the imaginary cylinder elongated along the axis of rotation and surrounding the inner core). So far the geomagnetic dipole between the reversals wanders inside of the cylinder, one can expect increase of the  virtual geomagnetic pole fluctuations relative to  the axis of rotation during the evolution of the Earth.

                            The other reason  is  existence of two mechanisms of convection simultaneously: the thermal convection, concerned with the radioactive heating, and compositional convection, which is   associated with the growth of the inner core. The latter mechanism is supposed to be  more efficient, because the heat is injected at the bottom of  the liquid core. Compositional convection  can  produce three times more energy than the thermal convection. In the thermal convection the radioactive heat sources are distributed smoothly in the bulk of the liquid core. It is quiet different from the compositional convection, where latent heat sources, concerned with crystallization process at the surface of the inner core, are localized at the inner core boundary. The shift of this boundary, where  maximum of the magnetic field generation is observed in 3D models,   relative to the observer at the surface of the planet, can change spatial spectrum of the observable magnetic field. 
      
        These more or less evident assumptions were already tested in some 3D models, however the obtained conclusions  still can not provide the robust scenario of the core evolution, supported by the paleomagnetic measurements  \cite{RP}. There are at least two reasons of such failure. 
        
          The first one is that geomagnetic field indeed does not reveal significant changes during the supposed period of the   inner core formation 1-2Ga. The other point is that 3D models, due to its complexity, provide too short time series, which are not sufficient for the evolutionary processes treatment. Moreover, having deal with the 3D models, one substitutes the black box of the MHD system in the core, with the another one, named the non-linear system of 3D partial  differential equations, which can be solved only numerically. Note that the parameters used in the models in its turn are very far from that ones in the core. Even if the model corresponds to observations, the level of understanding of the physics of such a complex  system is not satisfactory. Interpretation of the dynamo  process using the simpler and more obvious  scenarios is appreciated.
      
        It motivates us  to use further the simpler,  2D Parker's dynamo model    \cite{Parker1955}, which was  developed latter to the  mean-field dynamo theory  \cite{KR}. 
                     The energy sources: the $\alpha$-effect and differential rotation, are taken from 3D simulations, and can vary with change of the inner core's  size in the prescribed manner. It is  also considered the different spatial distributions of the $\alpha$-effect and differential rotation, which depend on the intensity of the heat sources in the core. As a result we present  dependence of the observed at the surface of the core magnetic field on the radius  of the solid core and analyze how the spatial spectrum of the magnetic field varies during the Earth's evolution. This analysis helps to distinguish the main features of the flows in the liquid core responsible for generation of the dominant dipole magnetic field in the past.

  \section{Dynamo model}

\label{section:2}
The mean magnetic field $\bf B$  is governed by the induction equation
\begin{equation}\label{Parker:1}
{\partial{\bf
B}\over\partial t}=\nabla \times \Big( \alpha\,{\bf B}+
{\bf V}\times {\bf B}
-\eta\, {\rm rot}{\bf B}   \Big), 
\end{equation}
where $\bf V$ is the large-scale velocity field,  $\alpha$ is the $\alpha$-effect,  and $\eta$ is a magnetic diffusion.
 The  magnetic field ${\bf B}=\left( {\bf B^p},\, {\bf B^t} \right)$
has two parts: the poloidal component ${\bf B^p}=\nabla\times {\bf A}$, 
where $\bf A$ is the  vector  potential of the magnetic field, and the toroidal component $\bf B^t$.

 In the axi-symmetrical case  the vector potential $\bf A$ and $\bf B^t$ have 
the only one azimuthal component
 in the spherical system of coordinates
$(r,\,\theta,\, \varphi)$: ${\bf A}(r,\,\theta)=(0,\, 0,\, A)$, and ${\bf B^t}(r,\,\theta)=(0,\, 0,\, B)$. 

The poloidal field  can be written in the form: 
\begin{equation}\label{Parker:2}
\displaystyle
{\bf B^p}=
\left(
 {1\over r\, \sin\theta}{\partial\over\partial \theta }\left( A\, \sin\theta \right),\,
-{1\over r} {\partial \over \partial r}  \left( r\, A \right),\, 0
 \right).
\end{equation}

In terms of scalars $A$ and $B$ Eq\ref{Parker:1} is reduced to the following system of equations:
\begin{equation}\label{Parker:3}
\begin{array}{l}
\displaystyle
{\partial{
A}\over\partial t}=\alpha {B} + \left({\bf V}\times\,{\bf B}\right)_\varphi
+\eta \left( \nabla^2 - {1\over r^2\sin^2\theta}  \right) {A}
\\  \\ 
\displaystyle
{\partial{B}\over\partial t}={\rm rot}_\varphi \left( \alpha\,{\bf B} +
 {\bf V}\times\,{\bf B}\right)+\eta
 \left( \nabla^2 -{1\over r^2\sin^2\theta}  \right){B},
\end{array}
\end{equation}
where  the subscript $\varphi$ corresponds to the azimuthal component of the vector, and $\eta$ is equal to unity.

Eqs(\ref{Parker:3}),   solved in the spherical shell
 $r_c\le r\le r_\circ$ with variable $r_c$, and $r_\circ=1$,
 are closed with the pseudo-vacuum boundary conditions: ${ B}=0$, and $\displaystyle {\partial \over\partial r} \left( r A\right)=0$ at $r_c$, and $r_\circ$. 
  The fields are vanishing at the  axis of rotation $\theta=0,\, \pi:$  $A=B=0$. The simplified form of the vacuum boundary condition  for $A$ is  well adopted in dynamo community, and presents a good approximation of the boundary with the non-conductive medium \cite{Jouve}. The reason why
 the vacuum boundary condition is  used at the inner core boundary is  concerned with the weak influence of the inner
 core on the reversals statistics of the magnetic field \cite{Wicht}.

   In the general case velocity  $\bf V$ is a three-dimensional vector, which depends on  $r$ and $\theta$. Further we consider only the effect of the differential rotation, concerned with the $\varphi$-component of $\bf V$, leaving the input of the meridional circulation  $(V_r,\,V_\theta)$ out of the scope of the paper.
   The amplitude of the azimuthal velocity component  $V_\varphi=\Omega s$, where $\Omega$ is the angular velocity of the fluid, and $s$ is the distance from the axis of rotation $\bf z$, is defined by constant $C_\omega$.

 The model is closed by  the  $\alpha$-quenching in the local algebraic form:
  \begin{equation}\label{Parker:4}\displaystyle
     \alpha=C_\alpha\,{\alpha_\circ  \over 1+E_m(r,\,\theta)},
 \end{equation}
where $E_m$ is the magnetic energy, and $C_\alpha$ is a constant.

     The system (\ref{Parker:3},\ref{Parker:4}) was solved using the $4^{th}$-order Runge-Kutta method, where   spatial derivatives in the r.h.s. were   approximated 
       by the second-order central-differences at the mesh grid $(r,\, \theta)$ $(101\times 101)$. 
     These algorithms resulted in  C++ object  oriented code with OpenMP for parallelization. The  post-processor graphic visualization was organized using the Python graphic library MatPlotlib. All simulations were done under Ubuntu  OS. 
   See the details of the benchmark
      \cite{Jouve} 
   in \cite{R14}.

        To demonstrate dependence of solution of Eqs(\ref{Parker:3},\ref{Parker:4}) on the different  parameters   the  MPI wrapper was used   to run 
    the main program at two  cluster supercomputers:  Lomonosov in Moscow State University and at the Joint Supercomputer Center of RAS. 
      The wrapper called the main program  with the fixed different values of parameters, like radius of the inner core $r_c$, and amplitudes of the  $\alpha$- and $\omega$-effects,     
 $C_\alpha$,  $C_\omega$, at the different processors and then gathered all the data at the end of simulations.

    \section{Spatial distribution of the fields in presence of the rapid rotation}
            In the first approximation  effect of rotation results in elongation of all the fields along the axis of rotation. The linear theory predicts that derivatives of the velocity and temperature fluctuation fields along the axis of rotation is five  orders of magnitude smaller than in the perpendicular plane
        \cite{Roberts}, \cite{Busse}.               
                             The turbulent effects decrease this difference, leaving  it however still substantial. This feature distinguishes the planetary dynamo from the dynamo in the galaxies and in the majority of the stars, where rotation is not so strong. It means that in the considered  2D model gradients of the prescribed $\alpha_\circ$ and $\Omega$ should also reflect this feature.
                           
                                   This point is very tricky, because  the mean-field approach is based on existence of the intermediate scale, concerned with the averaging of the turbulent fields. This averaging  leads to the opposite effect, concerned with  smoothing of the sharp gradients of the velocity and temperature fluctuations  fields, which are indeed observed in 3D dynamo models. As a result the difference between the derivatives of the fields in $s$- and $z$-directions in the cylindrical system of coordinate should substantially decrease.
                                   
                                    Having these arguments in mind, only the large-scale features of the flow, taken from 3D simulations, should be included in the mean-field dynamo models. For the angular velocity $\Omega$ it is dependence only on $s$-coordinate, so that for large $s$ $\Omega>0$ and for small $s$ $\Omega$ is negative, see, e.g.,  \cite{BS08}. As regards to the distribution of $\alpha$ it should change sign at the equator plane and concentrate near the Taylor cylinder, where cyclonic convection is localized   \cite{R12}. 

   We start from the more explored to the moment in the geodynamo regime with the present size of the solid core $r_c=0.35$, and introduce the following 
   proxies to the $\alpha_\circ$- and $\Omega$-distributions:
\begin{equation}\begin{array} {l}\displaystyle 
         \alpha_0^I=\widehat{C}_\alpha(1-erf(1.25|z|))e^{-S_c (s - 1.1r_c)^2}s\cos(\theta), \\ \\\displaystyle
         \Omega^I=   -\widehat{C}_\Omega \left(1+erf(8(s-r_c))\right) \cos\left({\pi(s-r_c)\over 1-r_c}\right),
 \end{array}\label{aom1}
\end{equation}             
 with $S_c=67$, $s=r \sin(\theta)$, and $z=r \cos(\theta)$. The positive constants  $\widehat{C}_\alpha$, $\widehat{C}_\Omega$ satisfy conditions: the maximal values of $|\alpha_0^I|$, $|\Omega^I|$ are equal to unity.
\begin{figure}[ht!]   
 \def \ss {6cm}
\captionsetup{width=.8\linewidth}   
        \vskip -.5cm
 \hskip 1.2cm \epsfig{figure=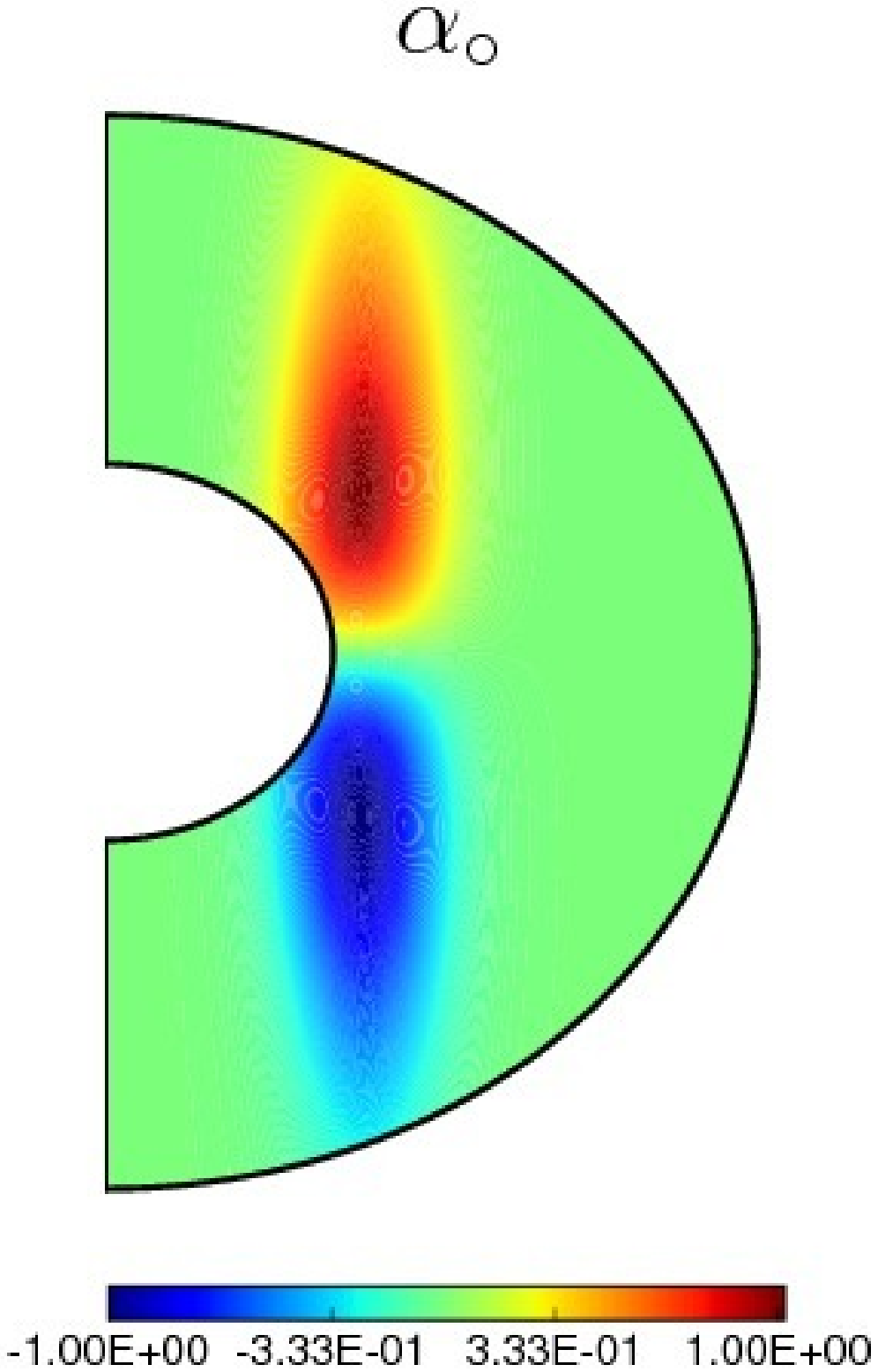,width=\ss}  \hskip 1cm \epsfig{figure=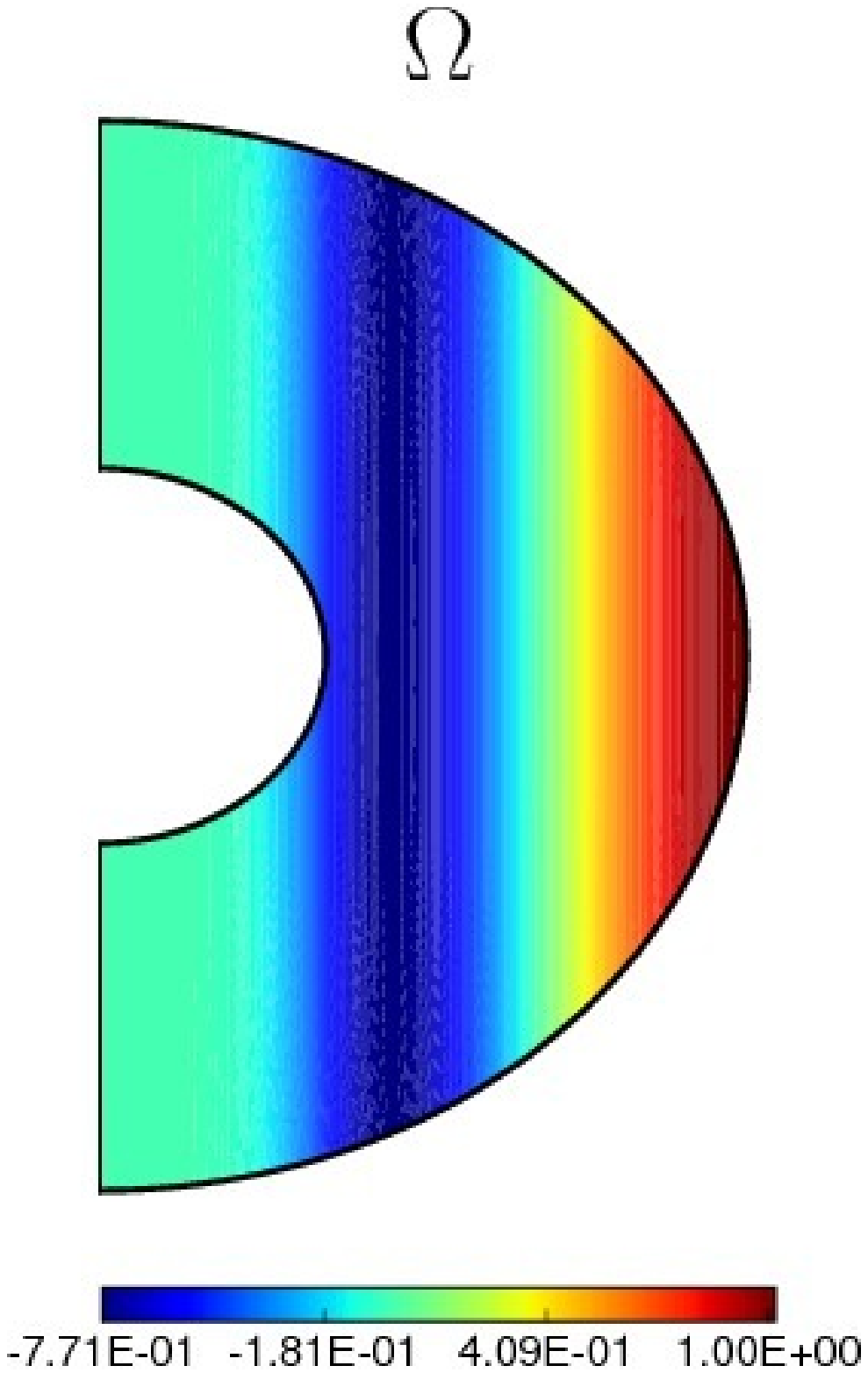,width=\ss}
  \vskip -1.8cm
  \caption{
    Meridional sections of $\alpha$-effect and angular velocity $\Omega$ for the small Rayleigh numbers. 
    } \label{1}
 \end{figure}
      These distributions, see Figure 1, correspond to the strongly geostrophic state near the threshold of convection generation (small Rayleigh number), where convection is concentrated outside of the Taylor cylinder. It appears   \cite{R14} that using these distributions one can  generate the Earth-like  magnetic field that   resembles the well-known $Z$-field distributions in Bragisnky's geodynamo model 
     \cite{Br75}.

    It is interesting that $B_r$-component is concentrated inside of the cylinder, see Figure 2, where $\alpha_\circ$ is small.  This is the essentially the non-linear effect, concerned with the small quenching effect inside of the cylinder and near the inner core boundary. The total magnetic energy $E_m$ with the main contribution from the toroidal field  counterpart (see distribution of $B$-component in Figure 2) is smaller inside of the cylinder, and equal to zero at the boundary due to the boundary conditions. Then, following  Eq.
(\ref{Parker:4}) one has 
          smaller $\alpha$-quenching in these regions.
          
            On the contrary, the large toroidal magnetic field outside of the cylinder sweeps out $\alpha$ from the region, suppressing the poloidal field generation.
            
               In other words the specific of our model is that the maximum of the poloidal magnetic field is determined by the intensity of the $\alpha$-quenching rather than by the amplitude of the original $\alpha_\circ$.

\begin{figure}[ht!]   
     \def \ss {6cm}   
\captionsetup{width=.8\linewidth}
       \vskip -.1cm
  \hskip 1.2cm \epsfig{figure=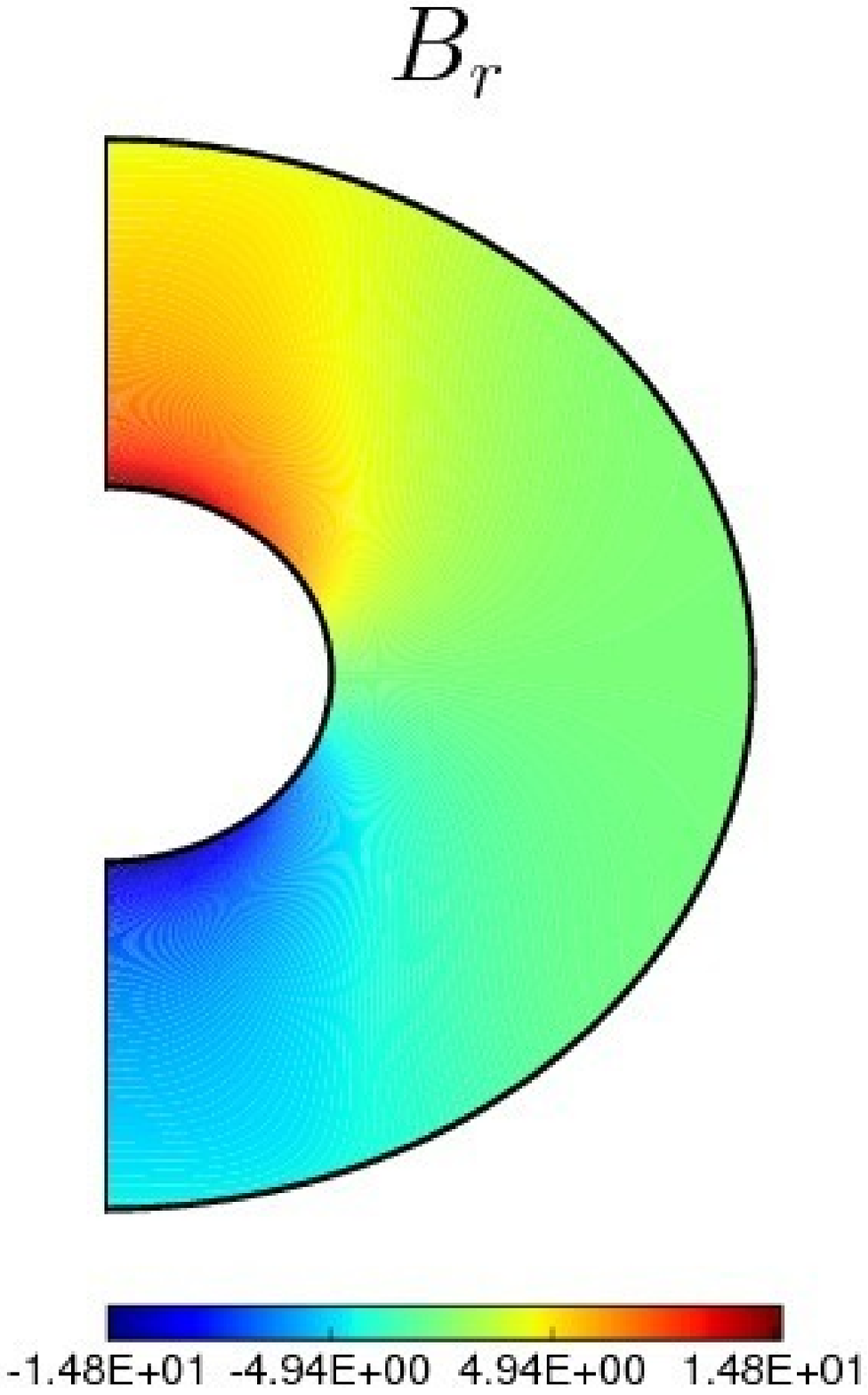,width=\ss}  \hskip 1cm \epsfig{figure=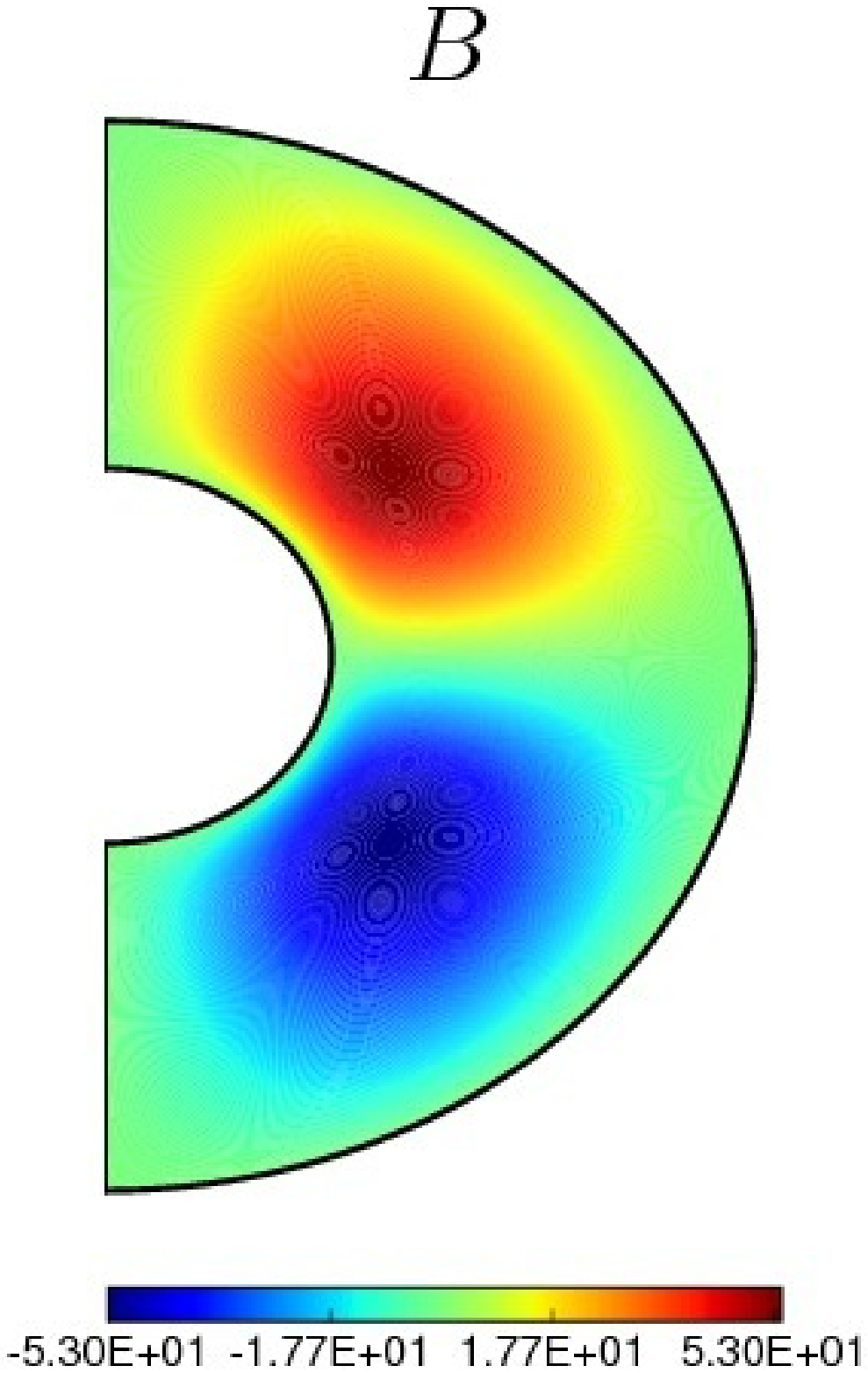,width=\ss}
        \vskip -1.8cm
          \caption{Distributions of the radial, $B_r$, and azimuthal, $B$, magnetic field components for the present time radius of the inner core, $r_c=0.35$.     } \label{2}
 \end{figure}
        If we believe that distributions of  $\alpha_\circ$ and $\Omega$  follow location of the Taylor cylinder then we can extrapolate distributions 
ef{aom1} to the smaller $r_c$. 
        However this approach leads to contradiction with the pillar of paleomagnetism that the geomagnetic field should be  dipole. This is demonstrated in Figure 3, where the gradual increase of the higher harmonics strength in the Mauersberger spectrum ${\cal S}(l)$ with decrease of $r_c$  is clearly observed.  
 \begin{figure}[ht!]   
    \def \ss {9.5cm}   
\captionsetup{width=.8\linewidth}
         \vskip -4.0cm
  \hskip 2cm \epsfig{figure=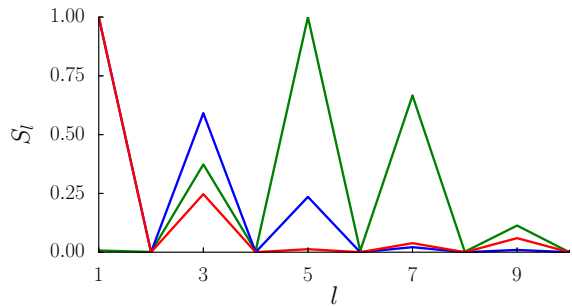,width=\ss}
      \vskip -4.2cm
       \caption{The normalized Mauersberger spectra for the different radius of the inner core: $r_c=0.15$ (green), $r_c=0.25$ (blue), $r_c=0.35$ (red). The maximal values of the original spectra before normalization  are $0.02,\, 0.56,\, 1.3$, correspondingly. The  flow depends on the inner core radius.
    } \label{3}
 \end{figure}

           The decrease of the dipole  is closely related to the geostro\-phic balance in the core. In presence of the geostro\-phic balance: the balance of the Coriois force and pressure gradient, the  velocity and temperature fields  variations are elongated along the axis of rotation. Moreover, the magnetic field also ``feels'' the geograpical poles, so that  the geomagnetic dipole, as already was mentioned, prefers to stay inside of the Taylor cylinder. The decrease of the radius of the Taylor cylinder decreases the scale of the magnetic field, concentrated inside of the cylinder. As a result the dipole contribution to the spectrum becomes smaller. This exactly what we observe in 
   Figure 3.
   
        It is worthy to note that this effect is stronger than  the opposite effect, concerned with the radial decay of the magnetic field: the smaller is the inner core,  the larger is the distance from the surface of the core to the inner core boundary, where the magnetic field is strong. Then the ratio of the dipole component to the other harmonics at the Earth's surface will increase with increase of the liquid core thickness. However, as we have just   shown,  the effect, concerned with geostrophy, is stronger than the effect, concerned with the radial decay of harmonics.

   There are two ways out of this situation. Of course,     
                it would be naive to expect  now  from the paleomagnetic community estimates of evolution of the Mauersberger spectrum on the geological times. That is why   we can not exclude the enforce of the higher harmonics in the spectrum in the past at all. 
          However rejection of the  dipole hypothesis of the paleo field 
                 leads to the principal impossibility of any useful for  theoreticians  mathematical description of the   fields spatial structure.  Then  it is more instructive to consider how our model can be modified to adjust the dipole hypothesis in the past.

               Firstly we have to note that we considered the differential rotation  which does not depend on $z$ at all. This flow corresponds to the regime with the  very small  Ekman numbers $\rm E$, which is indeed expected in the inner core: ${\rm E}~\sim 10^{-16}$. The increase of $\rm E$  breaks geostrophy and can lead to increase of the flowing up to the surface of the liquid core magnetic field. In  other words, dependence on $s$-coordinate will change to dependence on $r$.
                  This effect can be observed in 3D dynamo models with moderate Ekman numbers. Recall that in majority of the dynamo models $\rm E$ is in the range $10^{-6}\div 10^{-4}$ that is still quite far from the geophysical values in the liquid core. In this sense the prescribed geostrophical $\Omega$ in our 2D model can be  more realistic. It is important that for the compositional convection, where the heat flux, associated with crystallization,  is injected  at the inner core boundary. Then, in presence of the  geostrophic state, decrease of  $r_c$  leads to the decrease of the dipole field contribution. And on contrary, when the radial Archimedean  force is  quite strong in the thermal convection models, the magnetic field expands to the surface of the liquid core, increasing the scale. In this case the dipole field can be quite strong at the surface of the liquid core even at the  small $r_c$.
             
                          Situation is even worth if we take into account that increase of the heat sources leads to the shift of convection from the part of the liquid core outside of the Taylor cylinder to its inner part  \cite{GR}. If for the present size of the inner core such convection still generates  the dipole field, then for the smaller value of $r_c$ there is no chance for the dipole magnetic field. The same is for the thermal convection with the prescribed temperatures at the inner core and mantle boundaries, where  the density of the heat flux $q\sim 1/r$. For this model  generation of the magnetic field will be more effective  near the inner core boundary, where $q$  is larger.
                          
                              Summing up, we have that localization of the magnetic filed generation near the liquid core in presence of the geostro\-phic balance leads to decrease of the spatial scale of the magnetic field at the liquid core surface.
          It means that thermal convection with the radioactive heating suites  better   to the dipole filed generation in the past than the compositional convection, or the thermal convection  model with the fixed temperatures at the boundaries.
         
              We note that only taking into account of the inner core evolution allows   to come to such  a conclusion and  reject some models, which give similar configurations of the magnetic field at the surface of the Earth for the present value of the  inner core radius $r_c$.
              
  Besides the mentioned above  increase of the scale, produced by the radioactive heating, there is the another  
         physical effect, which leads to the same result. 
               The majority of the modern 3D dynamo models is based on the Boussinesq approximation of convection. It means that effect of compressibility  is taken into account only for derivation of the Archimedean forces. Then, the kinetic helicity, $\chi={\bf V}\cdot{\rm rot}{\bf V}$,  closely connected to the $\alpha$-effect, is generated near the boundaries of the liquid core, and in the vicinity of  the Taylor cylinder. For the realistic values of Ekman number the scales of the boundary layers, where helicity of the incompressible fluid is generated, are too small to generate the magnetic field. It means that extrapolation of helicity profiles to the realistic $\rm E$ can  lead to the break of the  magnetic field generation at all.
           
                However, in the compressible fluid kinetic helicity can generate in the bulk of the volume due to expansion/compression of the rotating flowing up/down fluid particle. This effect leads to increase of the $\alpha$-effect spatial scale. The drop of the density $\rho$ from the inner core boundary to the core-mantle boundary is 20\%. This is quiet enough for discernible contribution to the total $\alpha$-effect   \cite{R12}. Note that this effect has no connection to the inner core at all. In another words, we assume that for the state with the small inner core, where the drop of $\rho$ is even larger than for the present time,   effects of compressibility should be taken into account to provide generation of the large-scale magnetic field in Parker's dynamo model. 

               Fortunately, having deal with the mean-field dynamo model we can drop out some  details of the flow on the scales, including the inner core size in the past, leaving the geostrophy of the flow as the main feature.  
      Further we disconnect the localization of the $\alpha$-effect and differential rotation from the boundary of the  Taylor cylinder. Then, growth of the  inner core will lead only to the increase of the bulk of the core, but will not change substantially distributions of the energy sources in the model. Such assumptions are valid to the compressible fluid and thermal convection with distributed over the liquid core heat sources.
       
       To explore this possibility we consider the smoother distribution of $\alpha_\circ$, see Eqs(\ref{aom1}), with $S_c=17$, and fixed $r_c=0.35$, even if the volume of the liquid core in Eqs(\ref{Parker:3}) was changed. For the angular velocity $\Omega$ we used the following dependence 
          $\Omega^{II}= -C_\Omega\cos(\pi s)$, where we kept geostrophy, i.e. dependence on $s$ coordinate only, and the change of $\Omega$  sign from negative to positive with increase of $s$.

These assumptions let  to generate the dipole magnetic field, see Figure 4, which is larger  
 at $r_c=0.25$, than in the present time, as well as at the very small $r_c=0.15$.
 \begin{figure}[ht!]   
    \def \ss {9.5cm}   
\captionsetup{width=.8\linewidth}
      \vskip -4cm
   \hskip 4cm \epsfig{figure=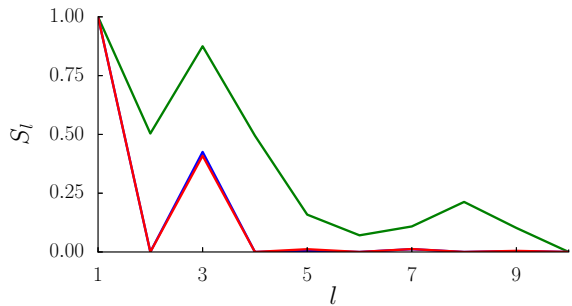,width=\ss}
      \vskip -4.2cm
      \caption{The normalized Mauersberger spectra for the different radius of the inner core: $r_c=0.15$ (green), $r_c=0.25$ (blue), $r_c=0.35$ (red). The maximal values of the original spectra before normalization  are $0.5,\, 2.9,\, 2.6$, correspondingly. The case with  the reduced dependency  on the inner core radius.
    } \label{4}
 \end{figure}
 As it follows, the ratio of the dipole field to the higher harmonics is the same for the range  
    $r_c=0.25\div0.35$, but for the smaller $r_c$ dipole is already comparable to  the   octupole, $l=3$. The total  magnetic field at the small $r_c$ is substantially weaker  than in the present time.
 
       In spite  of the fact that our distributions of $\alpha_\circ$ and $\Omega$ do not depend on $r_c$  explicitly, model feels the inner core due to the imposed vacuum boundary condition at the inner core boundary. As we already mentioned above, this trick leads to the effective increase of the poloidal magnetic field generation at the boundary, clearly observed in Figure 5. We can associate this effect with the release of the latent heat,   as well as with the enhanced viscous and Maxwell stresses at the inner core boundary in the more complex dynamo models.

 \begin{figure}[ht!]    
        \def \ss {4.62cm} 
\captionsetup{width=.8\linewidth}  
      \vskip -0.2cm
     \hskip -.2cm \epsfig{figure=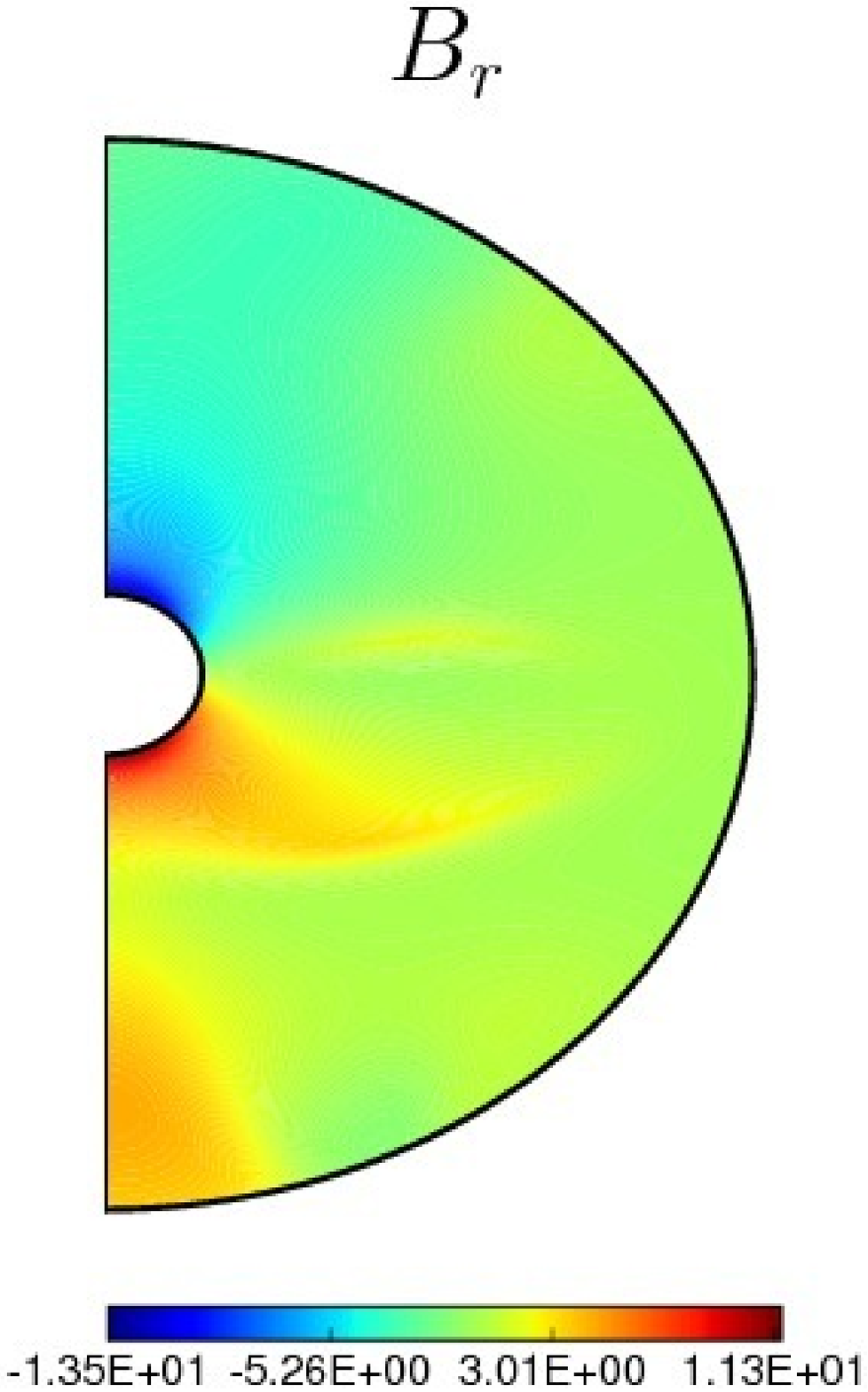,width=\ss}
     \hskip 1.3cm \epsfig{figure=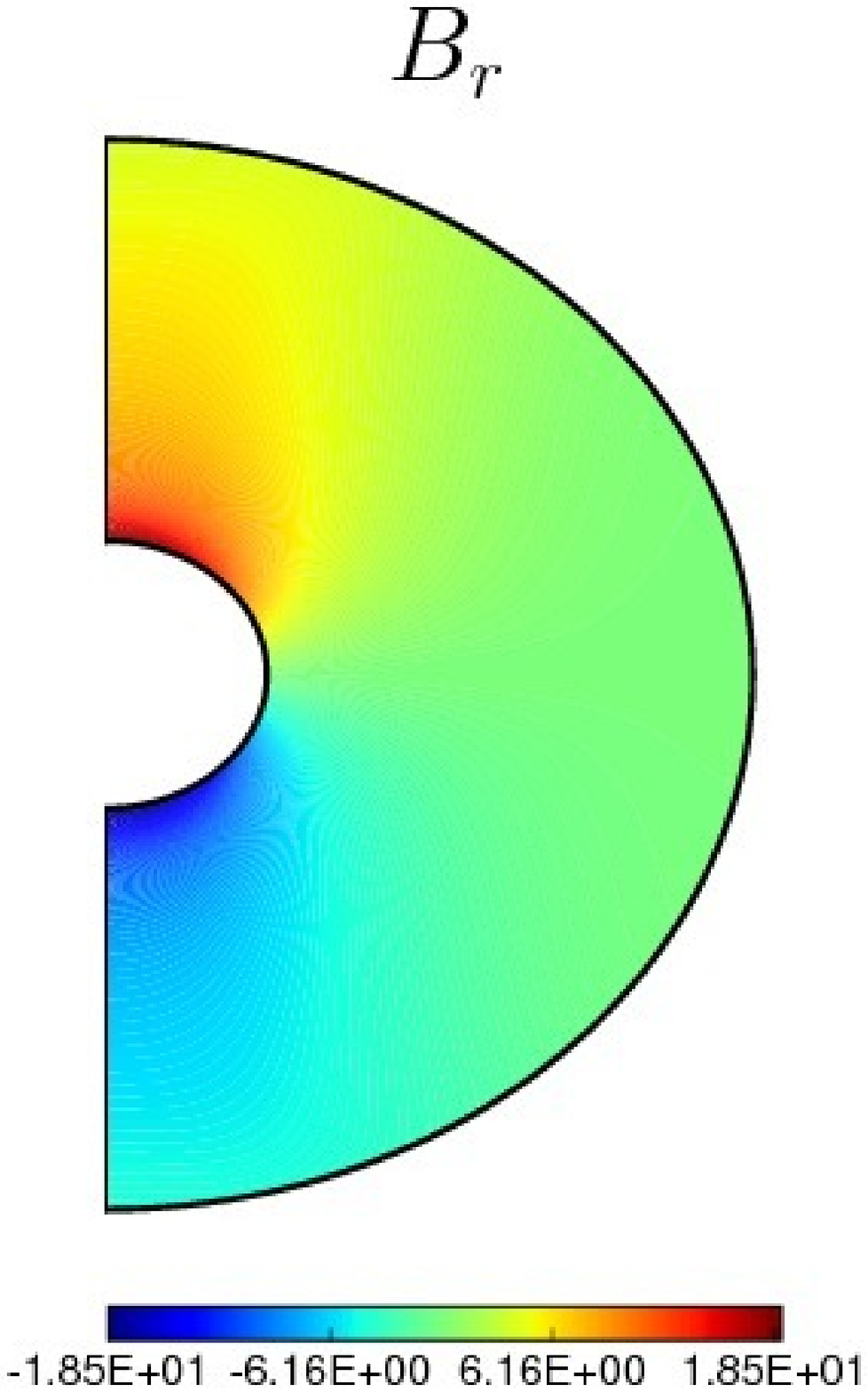,width=\ss}
              \hskip 1.3cm \epsfig{figure=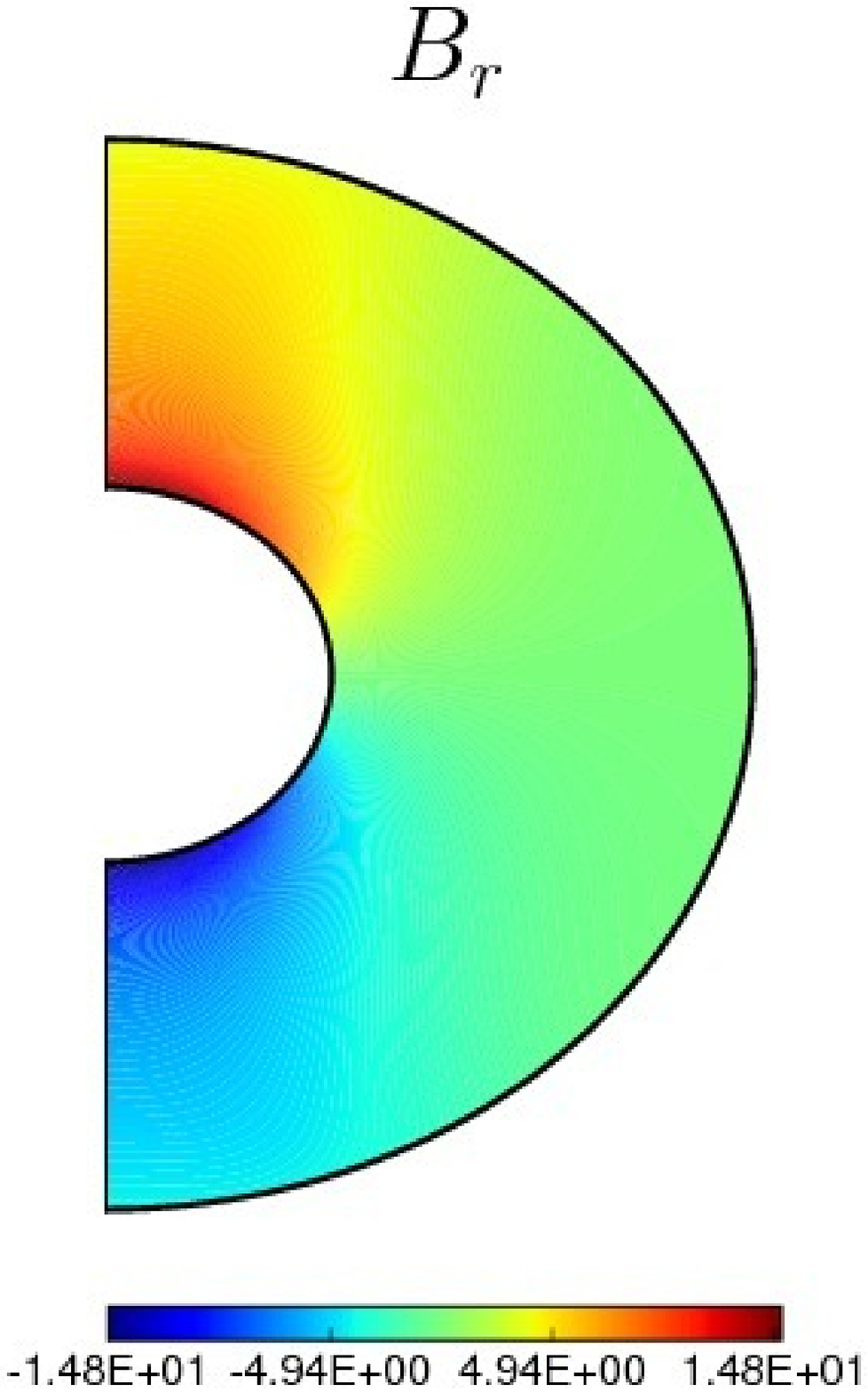,width=\ss}
        \vskip -1.5cm
       \caption{Distributions of the radial, $B_r$, component of the magnetic field for the inner core radius $r_c=0.15,\, 0.25,$ and $0.35$.
             } \label{5}
 \end{figure}
We conclude that using this technique we provided the  large-scale distributions of $\alpha_\circ$ and $\Omega$ in 
        Eqs(\ref{Parker:3})  took into account the growth of the inner core in the model, and obtained the dipole field configurations of the magnetic field for all $r_c$.

  \section{The random $\alpha$-effect}
      To the moment we did not discuss the time evolution of the magnetic fields, considering  its  spatial distributions only. As usually, the dipole magnetic fields in the  mean-field model with  geostrophic $\alpha_\circ$ and $\Omega$ 
                            are stationary. Increase of the energy sources leads to  the switch from the dipole magnetic field  to the  multi-polar state, with already fluctuating, and may be reversing, dipole. Even if we find the transition region in the phase space,  where the dipole's magnitude is still comparable to  the other harmonics, the  volume of this phase space will be  very small, and it would be difficult to justify correspondence of exactly these parameters to the geodynamo regime. In other words, the sharper is the boundary  between  two states, the less is the probability of switch between these states.
                         
                          Anyway, if we consider such fluctuating regimes, evolution of the magnetic dipole is very far from that one, observed in the geodynamo. The mean-field models demonstrate  oscillations of the dipole, which resemble superposition of the periodic harmonics. On contrary, paleomagnetic observations point at existence of two attractors of the magnetic dipole at the geographic poles, and the quick transitions of the dipole between attractors: the reversals. Between the reversals there is the so-called regime in oscillations, where the dipole wanders around the pole, without change of the sign.
   The 3D dynamo models can indeed produce similar to observable behavior of the magnetic dipole, see, e.g., review    \cite{RK2013}, but the mean-field models require additional modification. 
 
          The hint is that in 3D models there is the turbulence, which triggers the large-scale fluctuations. The mean-field dynamo models are too simple nonlinear systems to generate the small-scale fluctuations, and these fluctuations should be injected into the system by hand. It means that averaged  quantities, used in the mean-field dynamo models, like $\alpha$-effect and differential rotation, must fluctuate.

           Peter Hoyng was the first one who used these ideas in  the solar dynamo 
   \cite{Hoyng}, where  minima of the solar activity  are associated with the breaks of the dynamo cycle, caused by such fluctuations.
   Latter, stochastic $\alpha$-effect was used in the Galerkin's  dynamo models \cite{Sobko},
  where the geomagnetic dipole evolution was very similar to that one in the paleomagnetic records. The further analysis of the phase-space of 2D Parker's models,  using  the finite-differences   
      \cite{R16}, revealed some restrictions on the form of fluctuations of $\alpha$-effect. It looks attractive to use further the random $\alpha$-effect   and explore how the growth of the inner core effects  on the reversals statistic.
 To introduce fluctuations of $\alpha$-effect we modify Eq.~(\ref{Parker:4}) as follows:
  \begin{equation}\label{Parker:5}\displaystyle 
       \alpha=C_\alpha\,{\alpha_\circ\left( 1+\epsilon \right)  \over 1+E_m(r,\,\theta)},
  \end{equation}
      where $\epsilon$ is the random normal variable (the same for all the grid points) with zero mean value and standard deviation $\sigma$. The new fluctuation, introduced by $\epsilon$, was applied with the time step $\tau_f=0.1$. Then, after it,  during the time $\tau_f$  $\alpha$ depended on the magnetic energy $E_m$ only.

 \begin{figure}[ht!]   
         \def \ss {6cm}   
\captionsetup{width=.8\linewidth}
        \vskip -0.cm
      \hskip 1.6cm \epsfig{figure=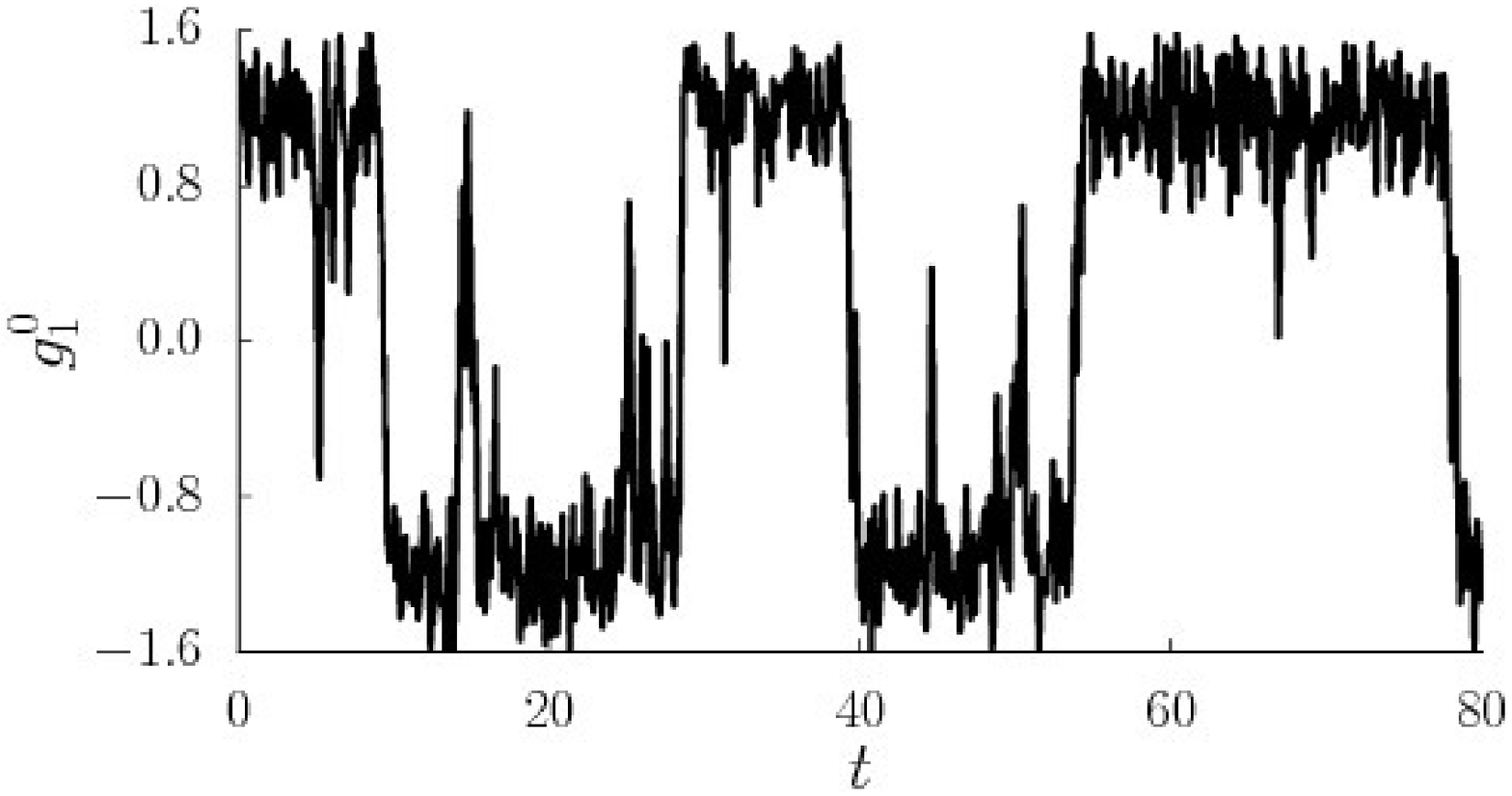,width=\ss}  
         \hskip 1.6cm \epsfig{figure=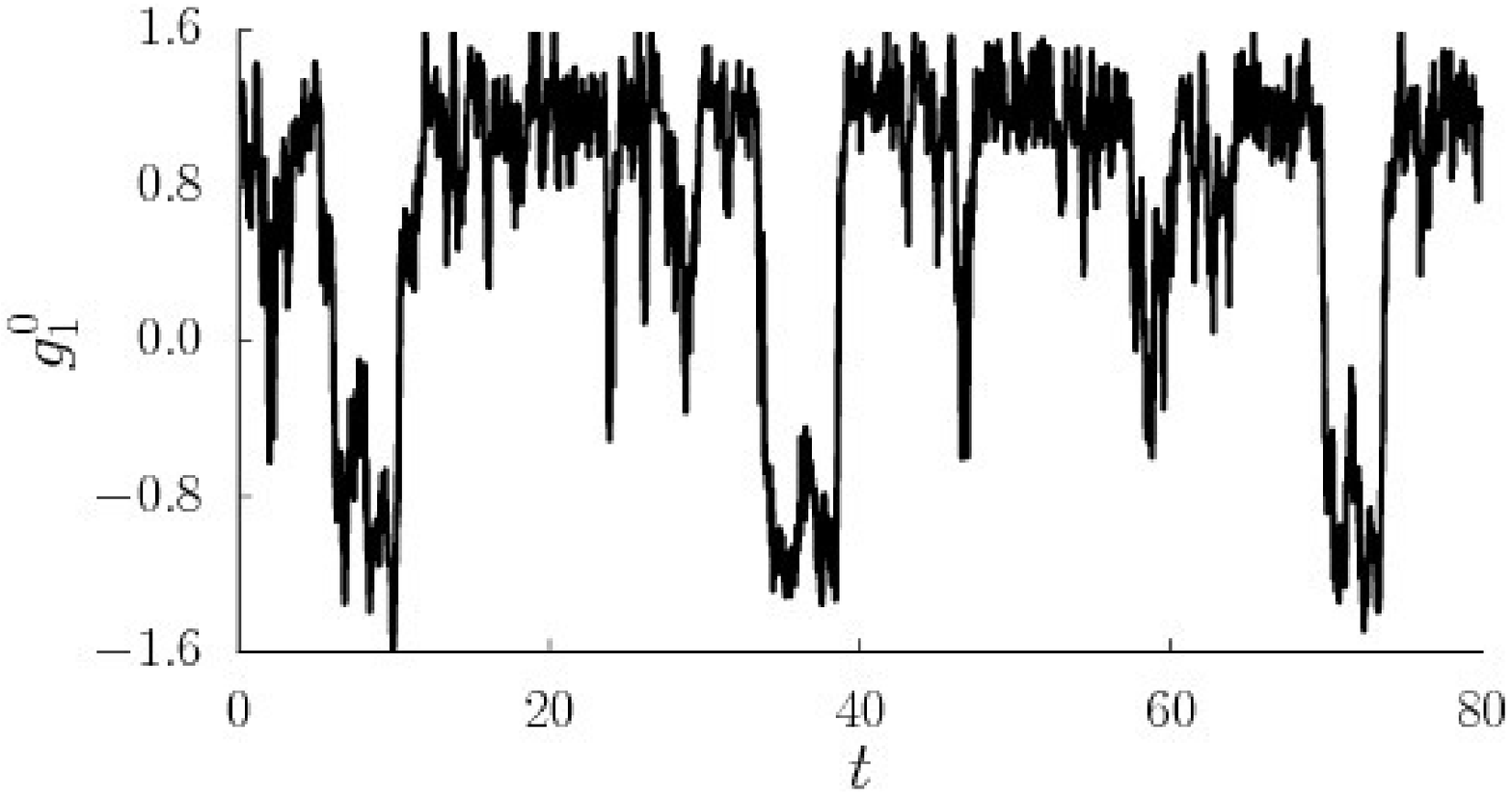,width=\ss}
\vskip -3.cm
       \hskip 5.6cm \epsfig{figure=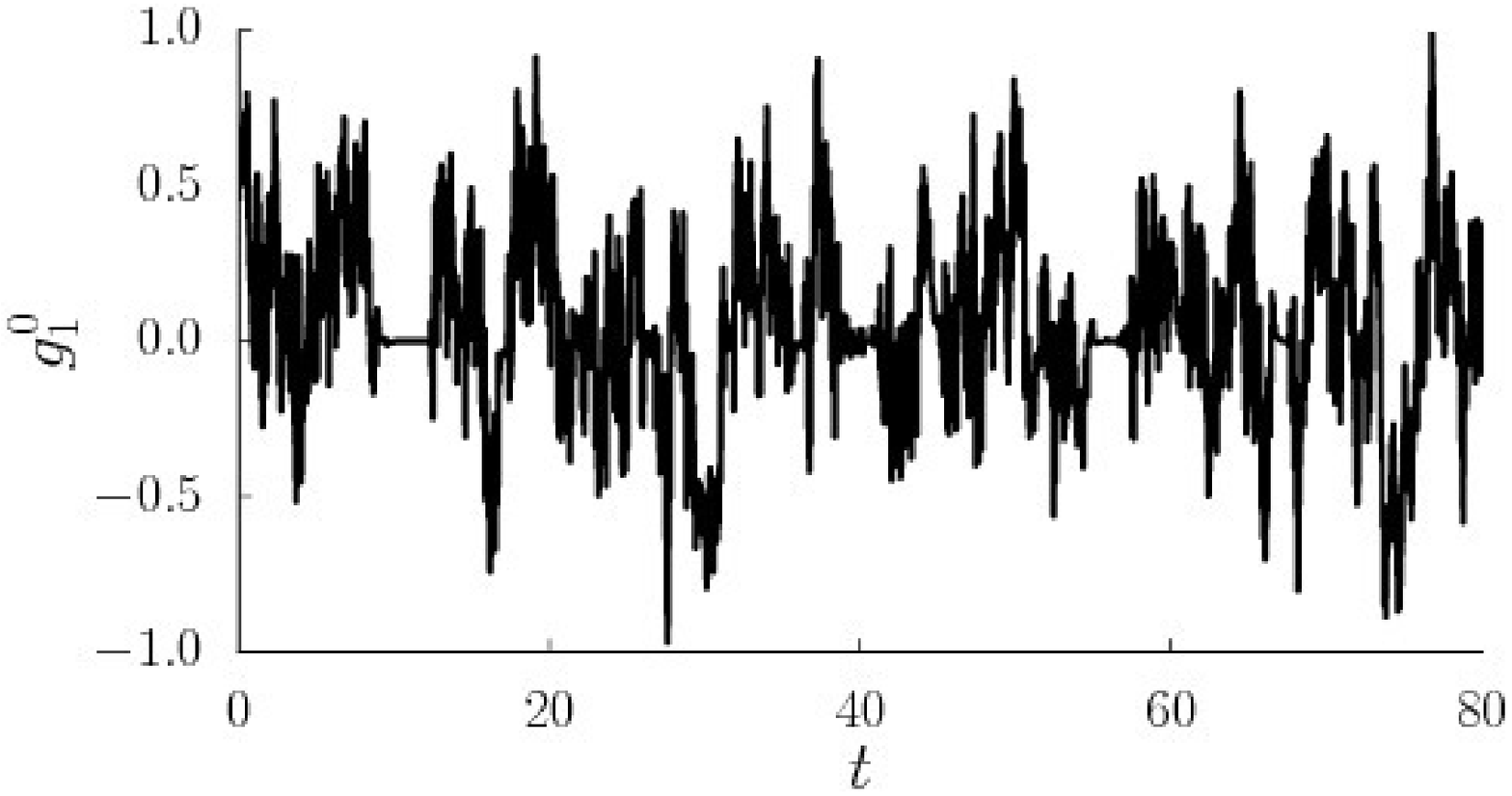,width=\ss}
             \vskip -2.2cm
            \caption{Evolution of the magnetic dipole $g_1^0$ with random $\alpha$ for $\sigma=0.7$, and different radius of the inner core:    $r_c=0.35$ (left),  $r_c=0.25$ (right),  $r_c=0.15$ (lower).
             } \label{6}
 \end{figure}

  We started from the state  with the core size  $r_c=0.35$ and the varied amplitudes   of the noise:   
              $\sigma=0.3,\,0.5,\,0.7,\,0.9$. The initial condition was taken from the solution, obtained in the Section 3. The first reversal was observed  at $\sigma=0.5$. The further  increase of $\sigma$ resulted in the gradual increase of the reversals number. For all $\sigma$ the magnetic field was dipole.

         Then we performed similar simulations for the other  two cases with $r_c=0.15$ and $0.25$. At the case with the small inner core, $r_c=0.15$, for all $\sigma$ the leading harmonic in the spatial spectrum was $l=3$ that corresponds to the  anti-symmetric  to the equator plane configuration, but with  the smaller scale than for the dipole field, $l=1$. 

We conclude that turbulent fluctuations in the liquid core lead to the decrease of the magnetic field scale at the surface of the core. Note that we considered only the dipole-like fluctuations of  $\alpha$ in  Eq.~(\ref{Parker:5}). Introduction of the independent   fluctuations at the every grid point, immediately leads to the pike in the spatial spectrum at the small scales even for $r_c=0.35$, see, \cite{R16}. This effect  is concerned with the absence of the inverse cascade of the magnetic energy in Parker's equations  Eq.~(\ref{Parker:3}).

                 The evolution of the magnetic dipole $g_1^0$ for $\sigma=0.7$ with varying $r_c$ is presented in Figure 6. For $r_c=0.25$ the amplitude of dipole fluctuations is slightly increased compared to the present time regime. However  the number of the reversals changed only from 5 to 6, the number of the  excursions  increased in more extent. In the both cases the  dipole component is the strongest one.  Obviously, the length of the time interval is too short to insist on any additional significant distinctions of the  cases with $r_c=0.25$ and $0.35$. 

                  The third case with $r_c=0.15$ is very different from the previous two. As was already mentioned, the dominating harmonic is $l=3$. As regards to the dipole's behavior, it spends more time in the low latitudes with the short  blowouts  to the poles. This behavior is natural to the state,  where  the harmonic  depends strongly on  interaction with the large number of the other harmonics. The memory on the attractors at the poles is lost. Thus,  fluctuations causes transition from the dipole magnetic field configuration to the octupole field at the small $r_c$. In other words, accordingly to this mean-field dynamo model,  the dipole configuration of the magnetic field in the past, if it did exist, was unstable.
                
\section{Conclusions}   
\label{section:5}
We tried to present the consistent scenario of  how information on the physical fields, obtained from the modern 3D dynamo simulations in the liquid core,
       could  be included to the  mean-field geodynamo model. This approach suggests that we distinguish the principal effects from the more complex models and then test them in the simpler ones. The correspondence of the obtained results to the 3D simulations and our expectations (based on some general knowledge on the system) let us to judge whether our suggestions were right or wrong.
       
        Following this way we come to the  quiet interesting phenomenon, concerned with the decrease of the magnetic field spatial scale in the past, caused by the small inner core. We emphasize that to test this effect in the 3D dynamo models, the low Ekman number regime is required. 
       
       Moreover, the majority of the dynamo models, which include sometimes the very different sources of the energy, 
  nevertheless, have the same size of the inner core (equal to the present core's size). The latter obstacle lets to generate the dipole magnetic field regardless to the physical mechanism of convection.
  
    On contrary, in the mean-field model, where the geostro\-phic balance is prescribed,  only for the comparable  size of the inner core to the modern one, the  magnetic field's spectrum  is dipole and stable to the turbulent perturbations.  The decrease of the inner core's size from $r_c=0.35$ to $0.15$ leads to the dipole magnetic field as well, which is however already unstable.
    
     It is worthy to note  that dipole field generation is supported by the ``volume'' effects, like compressibility of the liquid, homogeneously distributed radioactive heating sources, and can be violated by the processes  at the smaller scales: compositional convection at the stage of the small inner core, viscous and magnetic stresses  at the inner core boundary.
   
        So far the modern 3D dynamo models still lead to the  quite contradictory scenarios, we believe that  these suggestions will attract attention of the dynamo community and would  be carefully checked with the  higher level of accuracy, where it is possible.

\end{document}